\documentclass{ws-p9-75x6-50}
\usepackage{amssymb}

\begin{document}

\title{Geodesically complete cylindrical spacetimes}

\author{L. Fern\'andez-Jambrina}

\address{Departamento de Ense\~nanzas B\'asicas de la Ingenier\'{\i}a 
Naval, E.T.S.I. Navales\\ Arco de la Victoria s/n, E-28040-Madrid, Spain
\\E-mail: lfernandez@etsin.upm.es}

\maketitle

\abstracts{
A sufficient condition for an orthogonally transitive G2 cylindrical spacetime 
to be singularity-free is shown. The condition is general enough to comprise all
known geodesically complete perfect fluid cosmologies.}

\section{Introduction}

The main goal of this talk will be to discuss the emergence of 
singularity-free cosmological models in the frame of 
orthogonally-transitive $G_{2}$-spacetimes, more precisely in 
cylindrical spacetimes.

The issue was triggered in 1990 by Senovilla\cite{Seno} by means of 
publication of a diagonal $G_{2}$ spacetime with a radiation fluid as 
matter content. This spacetime had regular curvature invariants. The 
question arose then whether this spacetime was in fact 
singularity-free, that is, geodesically complete. By this it is meant 
that every causal geodesic can be defined from $-\infty$ to $\infty$ 
in the affine parametrization. In a subsequent paper\cite{Chinea} by Chinea, 
Senovilla and the author, the issue was settled. It was geodesically 
complete.

What about other spacetimes? Curvature invariants are easily checked 
and there is no problem in determining whether they are regular. 
Geodesics, on the other hand, are determined by a system of non-linear 
ODEs and it is therefore not easy to check whether they are complete. 
Since general relativity deals with lorentzian geometry, no riemannian 
metric is available and hence no Hopf-Rinow theorem relates geodesic 
completeness to metric completeness.

This talk is an attempt to simplify the issue in the context of 
cylindrical spacetimes, which has been the source of non-singular 
spacetimes with a perfect fluid as matter content.

We shall choose isotropic coordinates, $t,r$, in the timelike part of the 
metric, $z,\phi$ are cyclic coordinates corresponding to cylindrical 
symmetry,
\begin{equation}
ds^2=e^{2\,g(t,r)}\left\{-dt^2+dr^2\right\}+\rho^2(t,r)e^{2\,f(t,r)}d\phi^2
+e^{-2\,f(t,r)}\{dz+A(r,t)\,d\phi\}^2,\label{metric}
\end{equation}
and in order to ensure that the Riemann tensor is well defined, we 
shall impose that all metric functions are $C^2$ in their usual 
ranges, although geodesic equations just involve first derivatives of 
the metric. Furthermore, we shall require the axis to be locally 
flat\cite{Kramer}. Coordinates are chosen so that it is located at 
$r=0$.

Instead of writing the whole set of geodesic equations, we shall make 
use of the existence of constants of motion, $P,L$, related to 
translations along and rotations around the axis, and 
$\delta\in\{0,1\}$ for lightlike and timelike geodesics.

\section{Results}

The results can be summarized in the following theorem. 
Reference\cite{non} provides more details. A similar theorem is obtained for past causal geodesics changing 
$u$-derivatives for minus $v$-derivatives. This theorem comprises all known perfect-fluid regular spacetimes in 
the literature\cite{Diag}${}^,$\cite{Jerry}.

\begin{description}
\item[Theorem 1:] Under the previous requirements a metric has complete future causal geodesics if the following set
of conditions is fulfilled:

\begin{enumerate}
\item For large values of $t$ and increasing $r$, 
\begin{enumerate}
\item \label{Mxi1}$\left\{
\begin{array}{l}g_u\ge 0\\
h_u\ge 0\\
q_u\ge 0,\end{array}\right.$
\item \label{Mxi2} Either $\left\{
\begin{array}{lcl}{g_r}\ge 0&\textrm{or}& |g_r|\lesssim g_u\\h_r
\ge 0&\textrm{or}& |h_r|\lesssim h_u\\q_r\ge 0&\textrm{or}& |q_r|\lesssim
q_u.\end{array}\right.$
\end{enumerate}
\item For $L\neq0$ and large values of $t$ and decreasing $r$, 
\begin{enumerate}
\item \label{mxi1}$\delta\,g_v+P^2e^{2f}\,
q_v+\Lambda^2\frac{e^{-2f}}{\rho^2}h_v\ge 0$
\item \label{mxi2} Either $\delta g_r+P^2e^{2f}\,
q_r+\Lambda^2\frac{e^{-2f}}{\rho^2}\,h_r\le 0$ {or} $\delta g_r+P^2e^{2f}\,
q_r+\Lambda^2\frac{e^{-2f}}{\rho^2}\,h_r\lesssim \delta g_v+P^2e^{2f}\,
q_v+\Lambda^2\frac{e^{-2f}}{\rho^2}\,h_v.$
\end{enumerate}
\item \label{tt} For large values of the time coordinate  $t$, constants $a,b$ exist such that 
$\left.\begin{array}{c}2\,g(t,r)\\g(t,r)+f(t,r)+\ln\rho-\ln|\Lambda|\\
g(t,r)-f(t,r)\end{array}\right\}\ge-\ln|t+a|+b.$

\item \label{ax} The limit $\displaystyle\lim_{r\to 0}\frac{A}{\rho}$ exists.
\end{enumerate}

\end{description}
where $u,v$ are the usual ingoing and outgoing light coordinates, 
$q=g+f$, $h=g-f-\ln|L-PA|$.

\section*{Acknowledgments}
 The present work has been supported by Direcci\'on General de
Ense\~nanza Superior Project PB98-0772.

\end{document}